\documentclass[fleqn,10pt]{wlscirep}
\usepackage[utf8]{inputenc}
\usepackage[T1]{fontenc} 
\usepackage{hyperref}       
\usepackage{url}            
\usepackage{booktabs}       
\usepackage{amsfonts}
\usepackage{nicefrac}       
\usepackage{microtype}     
\usepackage{xcolor}   
\usepackage{graphicx}
\usepackage{footnote}
\usepackage{adjustbox}
\usepackage{pifont} 
\usepackage{subfigure}

\newcommand{\tick}{\textcolor{green}{\ding{51}}}
\newcommand{\cross}{\textcolor{red}{\ding{55}}}

\title{Large Scale MRI Collection and Segmentation of Cirrhotic Liver}

\author[1]{Debesh Jha}
\author[1]{Onkar Kishor Susladkar}
\author[1]{Vandan Gorade}
\author[1]{Elif Keles}
\author[1]{Matthew Antalek}
\author[2]{Deniz Seyithanoglu}
\author[2]{Timurhan Cebeci}
\author[1]{Halil Ertugrul Aktas}
\author[2]{Gulbiz Dagoglu Kartal}
\author[2]{Sabahattin Kaymakoglu}
\author[2]{Sukru Mehmet Erturk}
\author[1]{Yuri Velichko}
\author[1]{Daniela Ladner}
\author[1]{Amir A. Borhani}
\author[2]{Alpay Medetalibeyoglu}
\author[1]{Gorkem Durak}
\author[1,*]{Ulas Bagci}

\affil[1]{Machine \& Hybrid Intelligence Lab, Department of Radiology, Northwestern University, Chicago IL 60611, USA}
\affil[2]{Istanbul University, School of Medicine (Capa), Istanbul, Turkey.}

\affil[*]{Corresponding author: ulas.bagci@northwestern.edu}

\begin{abstract}
Liver cirrhosis represents the end stage of chronic liver disease, characterized by extensive fibrosis and nodular regeneration that significantly increases mortality risk. While magnetic resonance imaging (MRI) offers a non-invasive assessment, accurately segmenting cirrhotic livers presents substantial challenges due to morphological alterations and heterogeneous signal characteristics. Deep learning approaches show promise for automating these tasks, but progress has been limited by the absence of large-scale, annotated datasets. Here, we present CirrMRI600+, the first comprehensive dataset comprising 628 high-resolution abdominal MRI scans (310 T1-weighted and 318 T2-weighted sequences, totaling nearly 40,000 annotated slices) with expert-validated segmentation labels for cirrhotic livers. The dataset includes demographic information, clinical parameters, and histopathological validation where available. Additionally, we provide benchmark results from 11 state-of-the-art deep learning experiments to establish performance standards. CirrMRI600+ enables the development and validation of advanced computational methods for cirrhotic liver analysis, potentially accelerating progress toward automated Cirrhosis visual staging and personalized treatment planning.

\end{abstract}

\begin{document}

\flushbottom
\maketitle
%
%
\thispagestyle{empty}

\section*{Background and Summary}
\textbf{Clinical motivation.} Liver cirrhosis, the end stage of chronic liver disease (CLD), is a major global health concern. In 2019, it was the 11th most common cause of death, accounting for 2.4\% of global deaths \cite {GINES20211359,naturereview}. Viral hepatitis is currently the leading cause of end-stage liver disease but metabolic dysfunction-associated steatotic (fatty) liver disease (MASLD) is soon expected to become the top etiology due to the global increased rate of obesity and metabolic syndrome \cite{naturereview}. Other etiologies like alcoholic liver disease, autoimmune hepatitis and hereditary diseases are also significant contributors \cite{naturereview}. Cirrhosis is characterized by bridging fibrosis and regenerative nodules, leading to impaired liver function and eventually liver failure \cite{pellicoro}.  The accurate segmentation of cirrhotic livers from radiology scans enables clinicians to monitor the progression of liver cirrhosis over time, which is crucial for assessing disease severity and treatment response. Detailed segmentation provides precise information on the extent and location of liver damage, aiding in treatment planning such as liver transplantation and targeted therapies.

\textbf{Scarcity of MRI data despite its clinical importance.} 
MRI holds immense potential for diagnosing cirrhosis, offering superior soft tissue contrast for visualizing lesions and characterizing fibrosis. However, its adoption is hampered by data scarcity compared to CT. Unlike CT's standardized Hounsfield Units, MRI lacks a universal intensity scale, making generalizability for deep learning models a challenge \cite{hantze2024mrsegmentator}. Scanner and acquisition protocol variability, along with artifacts, voxel size variations, and registration errors, further complicate MRI data analysis \cite{litjens2017survey}. Despite these hurdles, MRI remains the preferred choice for long-term monitoring of chronic liver disease and hepatocellular carcinoma (HCC) detection \cite{ramalho2017magnetic}. By developing deep learning methods to overcome these challenges, we can unlock the full potential of MRI data and revolutionize cirrhosis diagnosis.

\textbf{Gold standard and alternative techniques.} Liver biopsy, the gold standard for assessing liver fibrosis severity, is an invasive procedure with potential risks \cite{GINES20211359}. This has driven the development of several non-invasive methods for assessing liver fibrosis. Laboratory-based indices utilize blood tests to estimate fibrosis level~\cite{Biopsy,NAFLD,prognostic}. Ultrasound-based elastography techniques measure liver stiffness using sound waves, providing an indirect assessment of fibrosis~\cite{Fibrotest,Attenuation}. Newer techniques like MR elastography (MRE) show promise for quantifying the degree of fibrosis. This method uses acoustic pressure waves to generate shear waves within the liver, allowing a more accurate assessment of liver stiffness compared to conventional MRI. However, MR elastography is expensive and not widely available. \cite{Lurie}. Despite the limitations, MRI is widely available and is the most valuable tool for liver assessment at the moment in detecting early-stage fibrosis due to the excellent soft tissue characterization~\cite{GINES20211359,Lurie}.

\textbf{Introducing CirrMRI600+, Addressing the Critical Need for Comprehensive MRI Data.} Deep learning-based assessment of radiological features of cirrhosis from MRI has the potential to provide an accurate, non-invasive method for determining the stage of liver fibrosis. This could eliminate the need for liver biopsy and assist clinicians in the early management of patients~\cite{lancet}. Accurately assessing cirrhosis severity from MRI relies heavily on two fundamental pillars: \textit{robust segmentation algorithms} (to assess liver volume) and \textit{high-quality, comprehensive datasets}. Segmentation is also crucial for subsequent deep learning models to effectively analyze the specific features within the cirrhotic region that hold diagnostic value. More importantly, despite the urgent need of MRI data,  the literature has a limited availability of MRI cirrhotic liver data; hence, development of such deep learning models is hindered. To meet this critical need for a large-scale dataset, we introduce CirrMRI600+, which consists of 628 high-resolution abdominal MRI scans from 339 patients with cirrhotic liver and their corresponding ground truth segmentation. The main contributions of our work are highlighted below: 

\begin{enumerate}
\item \textbf{Public release of CirrMRI600+:} We have developed and publicly released a novel dataset specifically designed for cirrhotic liver research. This dataset comprises 628 high-resolution abdominal MRI scans (310 T1-weighted (T1W) and 318 T2-weighted (T2W)) volumetric scans from 339 patients. Both contrast-enhanced and non-enhanced MRI scans are included, along with corresponding segmentation masks annotated by physicians.  CirrMRI600+ is a single-center,  multivendor, multiplanar and multiphase dataset. To the best of our knowledge, CirrMRI600+ is the first dataset specifically designed for liver cirrhosis research and incorporates both T1W and T2W MRI images.

\item{\textbf{Liver cirrhosis stage classification using MRI data:}} We classified liver cirrhosis into three stages based on radiological evaluations: mild, moderate, and severe—based on comprehensive radiological evaluations (please refer to the dataset (CSV file) for patient-wise grading).

\item \textbf{Benchmark evaluation:} We conducted benchmark evaluations for cirrhotic liver segmentation using 11 state-of-the-art (SOTA) algorithms on both the T1W and T2W MRI scans within CirrMRI600+. By making these benchmark results publicly available, we aimed to encourage the medical imaging community to develop robust segmentation algorithms for MRI analysis. These algorithms can potentially eliminate time-consuming steps like cirrhotic liver segmentation, ultimately leading to the development of efficient and clinically valuable tools. The comprehensive nature of CirrMRI600+, encompassing a large number of cirrhotic liver scans with diverse disease states and morphology, provides a strong foundation for training segmentation algorithms that can generalize well in the setting of advanced liver disease.

\item \textbf{Multimodal dataset:} 
The multimodal nature of CirrMRI600+ (including both T1W and T2W) offers a distinct advantage. T1W MRI excels at depicting anatomical structures, vascular structures and fat content, while T2W MRI provides superior soft tissue contrast. T1W images can reveal characteristic features like capsular retraction, while T2W images can highlight the heterogeneity of the cirrhotic parenchyma due to variations in fibrosis and fluid content. By incorporating both modalities, CirrMRI600+ empowers researchers to develop segmentation and classification (cirrhosis severity estimation) algorithms that leverage the complementary strengths of T1W and T2W imaging, ultimately leading to more robust and informative analysis of cirrhotic livers.
\end{enumerate}

\section*{Related Work}
The publicly available datasets for abdominal organ segmentation have traditionally been limited by data scarcity and a lack of organ diversity. While recent years have witnessed a positive shift towards increased data sharing, significant gaps remain. As shown in Table~\ref{tab:table1}, existing datasets predominantly cater to either single organs or multiple organs, with a clear bias towards CT scans. MRI datasets, while present, often lack a specific focus on cirrhosis. The Duke Liver Dataset, for instance, includes some cirrhotic cases, but its primary function revolves around liver segmentation and series classification tasks \cite{macdonald2023duke}. 

\begin{table*}
    \caption{Comparison of CirrMRI600+ dataset with other liver and abdominal organ segmentation datasets. ``--'' refers to missing information due to data unavailability. $\dagger$ Portion of patients are diagnosed with cirrhosis.}
    \footnotesize
    \centering
    \begin{tabular}{lccccccc}
    \toprule
    \textbf{Dataset} & \textbf{\# Organs} & \textbf{Cirrhosis} & \textbf{\# Patients} & \textbf{\# Slices} & \textbf{\# Anns per Scan} & \textbf{Modality} & \textbf{Year} \\
    \midrule

    \texttt{KiTS}~\cite{heller2019kits19} & 1 & \cross  & 300 & 23,337 & 997K & CT & 2019 \\ 
    \texttt{LiTS}~\cite{antonelli2022medical} & 1  & \cross & 201 & 29,402 & 3410K & CT & 2018 \\ 
    \texttt{MSD-Spleen}~\cite{antonelli2022medical} & 1 & \cross & 61 & 1,563 & 40K & CT & 2018 \\ 
    \texttt{MSD-Prostate}~\cite{antonelli2022medical} & 1  & \cross & 48 & 712 & 15K & MRI & 2018 \\ 
    \texttt{MSD-Pancreas}~\cite{antonelli2022medical} & 1  & \cross & 420 & 13,141 & 144K & CT & 2018 \\
    \texttt{Duke Liver}~\cite{macdonald2023duke} & 1 & $\dagger$  & 105  & 113,280 & -- & MRI  & 2023\\
    \midrule
    
    \texttt{BTCV}~\cite{landman2015multi} & 13  & \cross & 50 & 3,629 & 431K & CT & 2015 \\ 
    \texttt{VISCERAL}~\cite{jimenez2016cloud} & 20  &  \cross & 120 & -- & -- & CT \& MRI & 2016 \\ 
    \texttt{CHAOS}~\cite{kavur2021chaos} & 4  & \cross & 80 & 1,989 & 52K & CT \& MRI & 2019 \\ 
    \texttt{AbdomentCT-1K}~\cite{Ma-2021-AbdomenCT-1K} & 4  & \cross & 1112 & 34,497 & 3412K & CT & 2021 \\ 
    \texttt{AMOS}~\cite{ji2022amos} & 15  & \cross & 600 & 74,026 & 9952K & CT \& MRI & 2022 \\ 
    \texttt{TotalSegmentor}~\cite{wasserthal2023totalsegmentator} & 104  & \cross & 1228 & --  & --  & CT & 2023 \\ 
    \texttt{AbdomenAtlas-8K}~\cite{qu2024abdomenatlas} & 8  & \cross & 8,448 & 3,200K & -- & CT & 2024 \\
    \texttt{MRSegmentator}~\cite{hantze2024mrsegmentator} & 40  & \cross & 1200 & -- & --  & CT & 2024 \\
    \midrule
    \texttt{\textbf{CirrMRI600+}} \textbf{(Ours)} & 1  & \tick & 339 & 39,954 & 688K & MRI (T1W \& T2W) & 2024 \\ 
    \bottomrule
    \end{tabular}
    \label{tab:table1}
\end{table*}

\subsection*{Benchmarking for segmentation}
Significant advances have been made in the automatic segmentation of major anatomical structures from medical imaging. Notably, Wasserthal et al.\cite{wasserthal2023totalsegmentator} introduced TotalSegmentor, a comprehensive model capable of automatically delineating 104 distinct anatomical structures (27 organs, 59 bones, 10 muscles, and 8 vessels) in CT images.  This model was developed using an extensive dataset of 1204 clinical CT volumetric scans collected longitudinally from routine examinations. Implementing the nnU-Net architecture, TotalSegmentor achieves impressive accuracy exceeding 90\% on test datasets, with both the model and training data publicly accessible for academic research. Building upon this foundation, Hantze et al.\cite{hantze2024mrsegmentator} recently developed MRSegmentor to address the unique challenges of MRI segmentation. MRSegmentor was trained on a heterogeneous dataset comprising 1200 manually annotated MRI scans from the UK Biobank, 21 in-house MRI scans, and 1228 CT scans. While also utilizing the nnU-Net architecture, MRSegmentor demonstrates particular robustness in segmenting anatomically variable structures such as the liver and kidneys. However, the model exhibits limitations when segmenting smaller, more complex structures like adrenal glands and the portal/splenic vein. Similar to its predecessor, MRSegmentor and its associated dataset have been made publicly available to the academic community, facilitating further research and clinical applications in MRI-based anatomical segmentation.

\begin{figure*} [!t]
    \centering
    \includegraphics[trim=0.3cm 0cm 0cm 0cm, clip, width =0.8\textwidth]{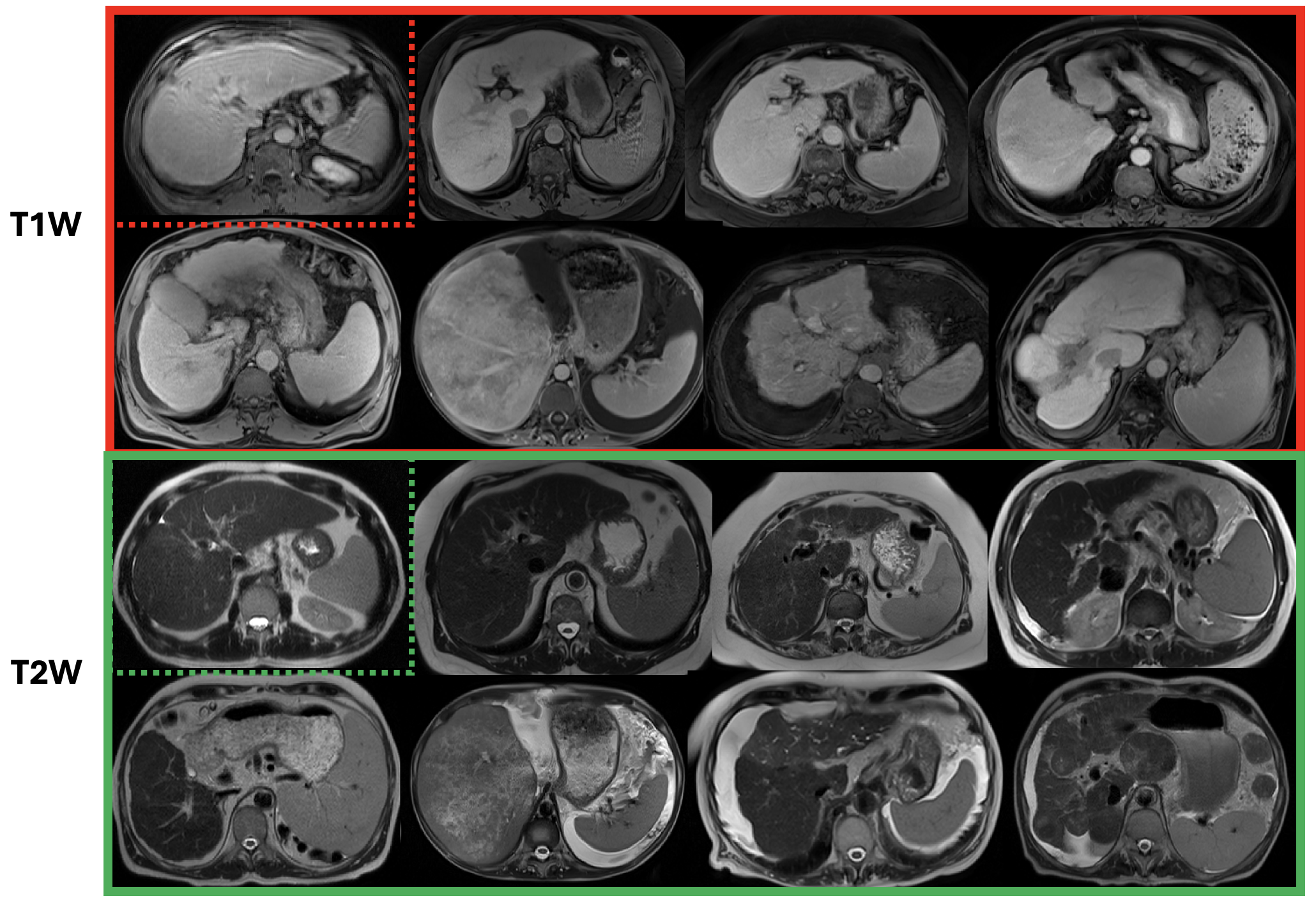}
        \caption{Different stages of fibrosis in liver MRI are shown in T1W and T2W samples, respectively. Varying levels of fibrotic tissues are observed across different scans, indicating the high variability and extreme challenge in texture and shape changes. To highlight the difference in fibrotic tissue diversity, MRI images in the dotted boxes (both red and green) are shown, they are having the smallest vulnerability to fibrotic tissue while still indicating cirrhosis. The first two rows are T1W, and the second two rows are T2W.}
    \label{fig:fibrosis_stages}
\end{figure*}

\section*{Method}
\subsection*{Ethical Approval} 
This retrospective study received approval from the Clinical Research Ethics Committee of the Istanbul Faculty of Medicine (approval date: 08/23/2024, protocol number: 2833959). Due to its retrospective nature, informed consent was waived. We adhered to strict patient privacy protocols, ensuring that all clinical and imaging data were fully anonymized. Ethical approval includes open publication of the data as well.

\paragraph{\textbf{MRI acquisition:}} 
We established a stringent protocol for MRI data acquisition with the following inclusion criteria: (1) Majority of patients were confirmed  with liver cirrhosis; 55 patients with normal abdominal findings or healthy livers were also included. (2) All imaging studies were required to meet quality assurance standards sufficient for expert radiologist annotation and review. (3) To ensure dataset heterogeneity, volumetric scans were acquired from three different scanner systems: Philips Achieva 1.5T, Philips Achieva 3T, and Siemens Symphony 1.5T, with comprehensive anonymization protocols implemented. The majority of T1W scans (>95\%) were post-contrast acquisitions obtained during the portal arterial phase to balance liver-to-vessel contrast. While scans with significant motion artifacts were excluded, the dataset does include cases with mild to moderate artifacts (e.g., motion, susceptibility) to reflect real-world clinical scenarios. (4) We prioritized the inclusion of diverse cirrhotic presentations across different etiologies and stages to capture the full spectrum of pathological variations and complications.
Following institutional review board approval and by standard clinical acquisition protocols, we retrospectively collected MRI data from 339 patients diagnosed with liver cirrhosis between 2015 and 2022. After quality assessment, the final dataset comprised 310 T1-weighted (T1W) and 318 T2W volumetric scans. Due to the retrospective nature of this study, the requirement for written informed consent was waived. The dataset poses minimal risk to patient privacy as all imaging data underwent rigorous de-identification, removing personal identifiers, including names, dates of birth, acquisition dates, and other directly identifying information. Technical metadata such as header size, image dimensions, pixel dimensions, data type, bits per pixel, voxel offset, and calibration parameters were preserved to maintain scientific utility. Diagnostic reports are available upon request, subject to additional confidentiality protocols.


\paragraph{\textbf{Patient Data:}} Our dataset exclusively comprises patients with liver cirrhosis, capturing real-world morphological alterations, including contour nodularity and hepatic segment atrophy/hypertrophy. This inherent complexity is crucial for developing robust, generalizable DL models. To enhance clinical representativeness, we included less common presentations such as parenchymal texture variations, focal liver lesions, and intrahepatic portal vein thromboses.
Figure~\ref{fig:fibrosis_stages} illustrates the diverse visual manifestations across cirrhosis stages in both T1W (top row) and T2W (bottom row) sequences. Notably, the leftmost images in each row show minimal visible fibrotic changes despite representing cirrhotic livers, highlighting the limitations of visual assessment in early-stage disease. This diagnostic challenge contributes to delayed detection, with many patients presenting only after developing decompensated cirrhosis or HCC~\cite{naturereview}—which is particularly concerning as morbidity and mortality rates increase with advancing fibrosis and progression from compensated to decompensated cirrhosis~\cite{sharma}.

\paragraph{\textbf{Properties of T1W and T2W:}} 
\textit{The T1W images} in our dataset were acquired using gradient echo (GRE) sequences with fat suppression. This combination, along with contrast enhancement, facilitated differentiation between T1W-bright structures resulting from fatty tissue versus vascular enhancement—features that would appear similar without fat suppression. These GRE sequences were acquired volumetrically (3D) with short repetition and echo times, yielding high-resolution images in three-dimensional space. Consequently, our T1W-based segmentation models were implemented as true 3D architectures. \textit{The T2W images} were predominantly obtained using accelerated techniques, primarily half-Fourier acquisition single-shot turbo spin-echo (HASTE). Unlike the T1W acquisitions, these sequences were acquired in a slice-by-slice fashion in the axial plane, generating pseudo-3D volumes. Due to the longer echo times necessary for T2W imaging, motion artifacts in the cranio-caudal plane from respiratory diaphragm movement were anticipated. To address this inherent limitation, our T2W-based segmentation models were trained using both volumetric (3D) and planar (2D) approaches.

\paragraph{\textbf{Data Standardization:}} 
We converted the large dataset of MRI scans from the DICOM format to the Neuroimaging Informatics Technology Initiative (NIFTI) format \url{https://nifti.nimh.nih.gov/} and uploaded it to the OSF server for public use~\cite{OSF}. NIFTI offers more efficient storage with comprehensive metadata, significantly reducing file size while improving dataset manageability. This conversion facilitated easier data sharing, enhanced reproducibility, and ensured compatibility with various analysis software. During conversion, all protected health information was removed from the DICOM files, and we verified the absence of duplicate images in the final dataset. 

\paragraph{\textbf{Segmentation Annotation:}} Our annotation process employed a semi-automated two-stage approach. First, we pre-segmented both T1W and T2W MRI scans using MRSegmentor~\cite{hantze2024mrsegmentator}. Second, four participating clinicians/radiologists refined these initial segmentations through manual annotation. The algorithm-generated masks required minimal corrections for early-stage cirrhosis cases where liver morphology closely resembled healthy livers. However, significant manual refinement was necessary for cases with advanced liver damage and associated anatomical alterations (splenomegaly, ascites, varices). This semi-automated workflow reduced the average annotation time from approximately 30 minutes to 10 minutes per scan, saving an estimated 207 radiologist working hours (equivalent to 26 working days). The refinement stage underwent multiple iterations until consensus was achieved among annotators. In total, we annotated 39,954 slices: 28,263 from CirrMRI600+$\rightarrow$T1W and 11,691 from CirrMRI600+$\rightarrow$T2W.

\paragraph{\textbf{Data split and evaluation metrics:}}
We split the dataset into training, validation, and test sets for CirrMRI600+$\rightarrow$T1W and CirrMRI600+$\rightarrow$T2W in an 80:10:10 split. This resulted in 248 cases for training, 31 cases for validation, and 31 cases for testing for CirrMRI600+$\rightarrow$T1W. Similarly, for CirrMRI600+$\rightarrow$T2W, we used 256 cases for training, 31 cases for validation, and 31 cases for testing. Although the split for T2W was not exactly 80:10:10,  we aimed to keep the distribution as close to the ratio as possible. Incorporating the domain shift caused by the device vendor, each split has variable scans. Nevertheless, we encourage researchers to select their training, validation, and test splits too. We evaluated liver segmentation performance using metrics such as the dice similarity coefficient (mDSC), mean intersection over union (mIoU), recall, precision, Hausdorff distance (HD95), and average symmetric surface distance (ASSD). Dice Similarity Coefficient and mIoU measure the overall volumetric overlap between predicted and ground truth segmentations. In clinical practice, these metrics correlate with the accuracy of liver volume estimation, which is crucial for surgical planning, transplantation assessment, and monitoring disease progression. However, these overlap metrics may not fully capture errors in anatomically significant regions (such as the porta hepatis) that occupy a small volume but have high clinical relevance. HD95 and ASSD assess the boundary accuracy of segmentations. HD95 represents the 95th percentile of the maximum distance between segmentation boundaries, effectively capturing the most significant localized errors while being robust to outliers. ASSD provides the average distance between boundaries, offering a more global assessment of boundary accuracy. In cirrhotic livers, these boundary metrics are particularly important for accurately capturing nodular surface changes and segment atrophy/hypertrophy that characterize advancing disease. However, these metrics may not distinguish between errors in clinically significant boundaries (such as those adjacent to major vessels) versus less critical regions.

\begin{table*}[t!]
\caption{Comparative benchmark of SOTA 3D segmentation networks across various metrics on \textbf{CirrMRI600+ $\rightarrow$T1} Liver Cirrhosis MRI dataset. Bold shows the best performance while red is the second-best.}
\centering
\footnotesize
\begin{adjustbox}{max width=\textwidth}
\begin{tabular}{lccccccc}
\toprule
\textbf{Method} & \textbf{mIoU} & \textbf{Dice} & \textbf{HD95} & \textbf{Precision} & \textbf{Recall} & \textbf{ASSD}   \\ 
\midrule

\texttt{VNet}~[\cite{milletari2016v}]  & 71.19 & 72.89 & 33.01 & 70.01 & 71.02 & 6.21  \\ 
\texttt{Attention UNet}~[\cite{oktay2018attention}] & 79.01 & 85.11 & 29.97 & 81.81 & \textcolor{red}{89.58} & 4.97  \\ 
\texttt{SynergyVNet3D}~[\cite{gorade2024synergynet}] & 76.11 & 78.77 & 27.55 & 85.12 & 86.72 & 5.34  \\ 
\texttt{TransBTS}~[\cite{wenxuan2021transbts}] & 63.42 & 76.11 & 36.92 & 74.84 & 84.01 & 7.39   \\ 
\texttt{UXNet3D}~[\cite{lee20223d}] & 77.60 & \textcolor{red}{86.58} & 30.45 & 83.25 & \textbf{91.09} & 4.76 \\ 
\texttt{TransUNet3D}~[\cite{chen2021transunet}] & 79.19 & 80.92 & 31.09 & 80.01 & 79.91 & 5.92  \\ 
\texttt{LinTransUnet}~[\cite{zhang2022dynamic}] & \textcolor{red}{83.77} &  86.11 &  25.77 &  86.99 &  85.79 &  \textbf{4.00}   \\
\texttt{SwinUNeTr}~[\cite{hatamizadeh2021swin}] & 81.02 & 82.01 & 30.66 & 81.32 & 80.97 & 5.01   \\ 
\texttt{nnUNet3D}~[\cite{zhou2023nnformer}]  & 82.22 & 85.72 & 26.78 & 86.67 & 85.98 & 4.38   \\ 
\texttt{nnFormer3D}~[\cite{zhou2023nnformer}]  & 83.03 & 86.09 & \textcolor{red}{25.18} & \textcolor{red}{87.11} & 85.72 & \textcolor{red}{4.01}   \\  
\texttt{nnSynergyNet3D}~[\cite{gorade2024synergynet}] & \textbf{84.51} & \textbf{87.89} & \textbf{21.04} & \textbf{88.72} & 87.76 & \textcolor{red}{4.01}  \\
\bottomrule
\end{tabular}
\end{adjustbox}
\label{table:T1segmentation_methods}
\end{table*}

\section*{Data Record} 
All data records collected in this study can be found in the OSF servers~\cite{OSF} (DOI:10.17605/OSF.IO/CUK24). All the medical images (MRI T1W and T2W) are stored using digital imaging techniques in the NIfTI format. The segmented images of the liver and the matched original medical images were stored in the NIfTI format after segmentation. The CirrMRI600+ dataset consists of 628 abdominal MRI scans (310 T1W and 318 T2W) collected from patients with liver cirrhosis. The dataset includes manual segmentations of cirrhotic livers across multiple stages of disease progression. All images were acquired in clinical settings and have undergone full anonymization. Our data collection adhered to the following protocol: \textbf{(1) Scanner diversity:} MRI scans were obtained from three different scanner models to ensure dataset heterogeneity: Philips Achieva 1.5T, Philips Achieva 3T, and Siemens Symphony 1.5T. (2) \textbf{Image Selection Criteria:} All included images met quality standards suitable for radiological assessment. Images with poor quality or significant motion artifacts were excluded, though the dataset does contain scans with mild to moderate artifacts (motion or susceptibility) to represent real-world clinical conditions. \textbf{(3) Contrast Enhancement:} Approximately 95\% of T1W images were acquired during the post-contrast portal venous phase to enhance organ-to-vessel contrast.
\textbf{(4) Patient Population: }The dataset includes images from patients diagnosed with liver cirrhosis exhibiting various morphological alterations, including: Contour nodularity, Hepatic segment atrophy or hypertrophy, Complications such as ascites, varices, and splenomegaly. \textbf{(5) Control Group:} A smaller set of non-cirrhotic control subjects (n=55) is included for comparison purposes.

The CirrMRI600+ dataset includes several metadata files to facilitate data usability and provide relevant subject information. The primary file, "\textit{...CompleteData-age-gender-evaluation.csv,}" contains demographic information (age, gender) and radiological (visual) evaluations of cirrhosis (1: Mild, 2: Moderate, 3:Severe) for all 337 patients included in the study. For researchers focusing specifically on T1W or T2W imaging, we provide separate files: "\textit{T1-age-gender-evaluation.csv}" details the demographic and evaluation data for the 310 patients with "\textit{T1-weighted MRI scans}", while "\textit{T2-age-gender-evaluation.csv}" provides the same information for the 318 patients with T2W scans. To support paired-image analysis, the "\textit{...-Paired-age-gender-evaluation.csv}" file identifies the 291 patients who have both T1W and T2W scans available. Additionally, the dataset includes "\textit{Healthy-demographics.csv}," which documents age and gender information for the 55 individuals in the control group. To ensure proper interpretation of all variables, we provide a "\textit{Labels.txt}" file that defines all column headings, coding schemes, and measurement units used throughout the metadata files.


\section*{Technical Validation} 
\subsection*{Validation of Data Collection.}
All images underwent quality assessment by participating radiologists prior to inclusion. Manual segmentations were performed by experienced radiologists following a standardized protocol as mentioned earlier in Methods section. The annotations provide ground truth segmentations of cirrhotic livers for all included MRI scans. We also provided radiological evaluation of each patient's MRI scans for cirrhosis severity (1: Mild, 2: Moderate, 3: Severe). This evaluation was done by our participating radiologist(s) based on organ volume and shape, decompensation and complication status (ascites, splenomegaly, HCC, varices), surface nodularity, and parenchymal heterogeneity-enhancement.

\subsection*{Validation of baseline segmentation models.}
We carefully selected several baseline methods from CNN-family (e.g., VNet~\cite{milletari2016v}, AttentionU-Net~\cite{oktay2018attention}) and from transformer-based methods (for example, nnUNet~\cite{isensee2021nnu}, Swin UNetR~\cite{hatamizadeh2021swin}, nnFormer3D~\cite{zhou2023nnformer}, LinTransUNet~\cite{zhang2022dynamic} and TransUNet3D~\cite{chen2021transunet}. All models were implemented in PyTorch 2.2.2 with CUDA 11.2 on dual Nvidia A6000 GPUs (48GB each) using PyTorch's Distributed Data Parallel. For 3D models, we employed a BCE-Dice loss function with AdamW optimizer, initial learning rate of 0.0001 with Cosine Annealing Scheduler, and decay rate of 0.001 every 10 epochs. Each volume was standardized to $256\times256\times80$ dimensions and processed with a batch size of 4. The 2D models maintained similar optimization parameters but used a batch size of 16 with the Adam optimizer. All networks were trained for up to 500 epochs with early stopping (patience=50) to prevent overfitting. Model performance was evaluated using multiple metrics: mIoU, Dice coefficient, HD95, precision, recall, and ASDD. To avoid overfitting and provide smooth training, we used rotation, translation, scaling, shear, and intensity transformation (shift, contrast adjustment, noise addition) to enhance the robustness and generalizability of segmentation methods. 


\color{black}
\textbf{Technical Validation of Segmentation Masks on CirrMRI600+$\rightarrow$T1W:} Table~\ref{table:T1segmentation_methods} presents a comprehensive evaluation of 11 SOTA 3D segmentation networks on the CirrMRI600+$\rightarrow$T1W dataset. As summarized, \textit{nnSynergyNet3D} achieved the highest overall performance with mIoU of 84.51, DSC of 87.89\%, HD95  of 21.04 mm, and precision of 88.72\%.  \textit{nnSynergyNet3D} performed better because of its synergistic and auto-configured continuous and discrete representation, allowing the model to capture fine and coarse features along with long-range dependencies due to its Transformer-inspired design. \textit{LinTransUnet} and \textit{nnFormer3D} demonstrated comparable performance in capturing cirrhotic liver tissue and its boundaries. Their performances were attributed to their Transformer based design with auto configuration, which enabled the models to learn and adapt to the liver's varying shape and complex boundaries. Again, this highlighted the importance of long-range dependencies. Conversely, \textit{nnUNet3D} demonstrated slightly poorer performance, underscoring the significance of Transformer-based representations for cirrhotic liver segmentation. Models like \textit{SwinUNeTr, TransBTS,} and \textit{TransUNet3D} do not significantly surpass CNN-based models such as \textit{nnUNet} and \textit{SynergyVNet3D}, showing the importance of auto-configuration and hybrid CNN-Transformer-based models.

\textbf{Technical Validation of Segmentation Masks on  CirrMRI600+$\rightarrow$T2W:} Table~\ref{table:T2segmentation_methods} presents evaluation of SOTA 3D segmentation networks on the CirrMRI600+$\rightarrow$T2W dataset. We observed similar results to those in the T1W segmentation results. \textit{nnSynergyNet3D} has a superior DSC value of 86.51\%, the lowest HD of 24.19 mm, and the lowest ASDD value of 3.96 mm. \textit{nnFormer3D} and \textit{nnUNet3D} are the other two competitive networks. The models such as \textit{SwinUNeTr, TransBTS}, and \textit{TransUNet3D} do not significantly surpass CNN-based models such as \textit{nnUNet} and\textit{ SynergyVNet3D}, emphasizing the importance of auto-configured and hybrid CNN-Transformer-based models for achieving competitive segmentation results.
It should be also note that the T2W images were predominantly acquired using accelerated techniques such as half-fourier single shot turbo-spin-echo (HASTE). The T2W images were acquired in a slice-by-slice fashion in the axial plane (the physics of HASTE generates pseudo-3D volumes). Due to the long echo times required for T2W image acquisition, motion artifacts in the craniocaudal plane from diaphragm movement during respiration are expected. Due to this, we trained T2W-based segmentation models using both volumetric data and 2D planar data. Additionally, we show the comparisons between the different models. By including the experiments on the 2D dataset, we want to develop a robust algorithm that could work well with motion artifacts that might have affected the 3D volumetric scans. We believe that for segmentation models using T2W images, analysis and segmentation using 2D images is an appropriate choice given the expected motion artifacts arising from the acquisition techniques.

\begin{figure*}  [!t]
    \centering
    \includegraphics[width=0.8\textwidth]{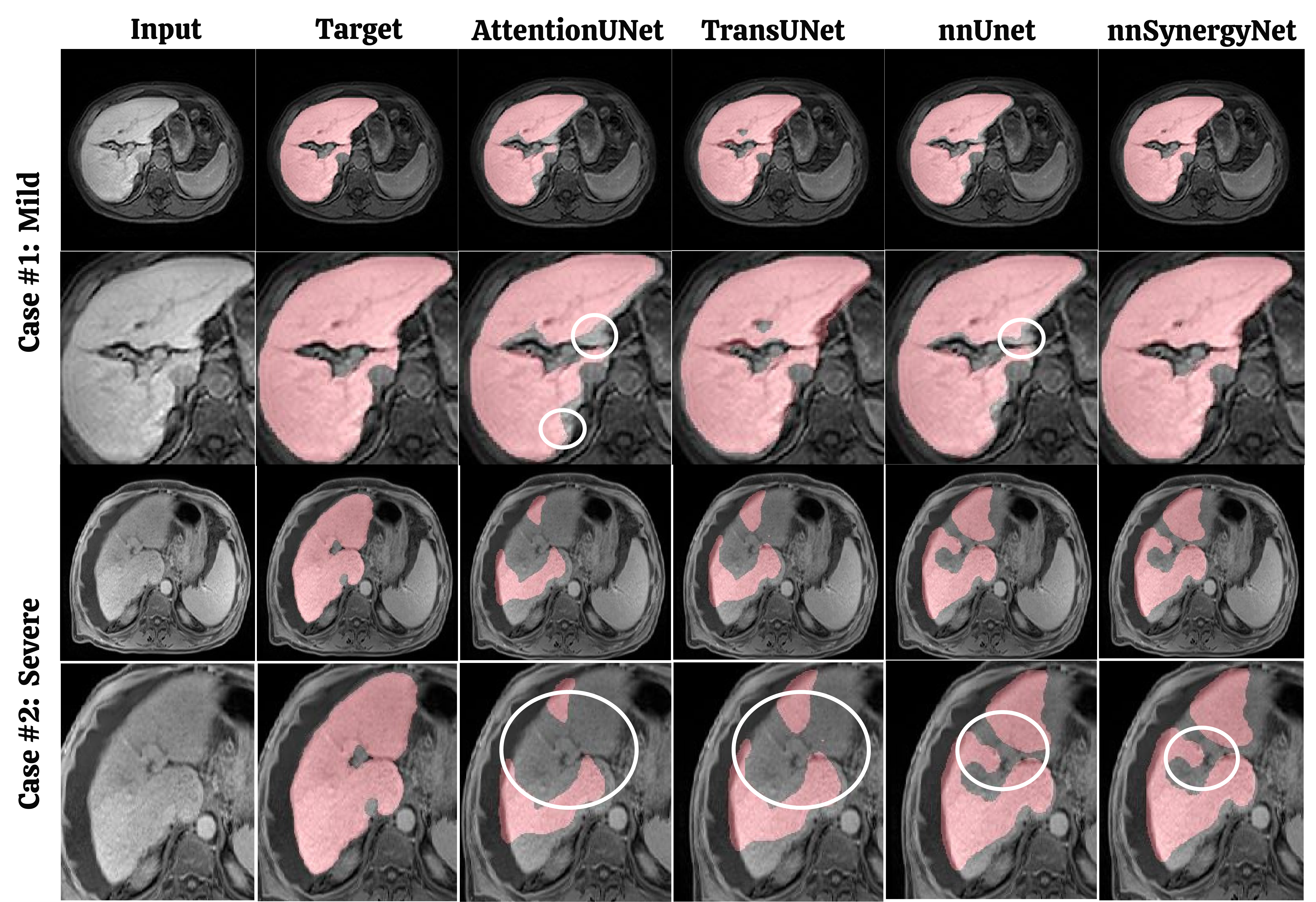}
    \caption{Qualitative results of different models on segmenting mild and severe cirrhosis from abdominal T1W MRI scans. The white bounding circles show major errors made by the models.}
    \label{qualitative-t1}
\end{figure*}

\begin{figure*}  [!t]
    \centering
    \includegraphics[width=0.8\textwidth]{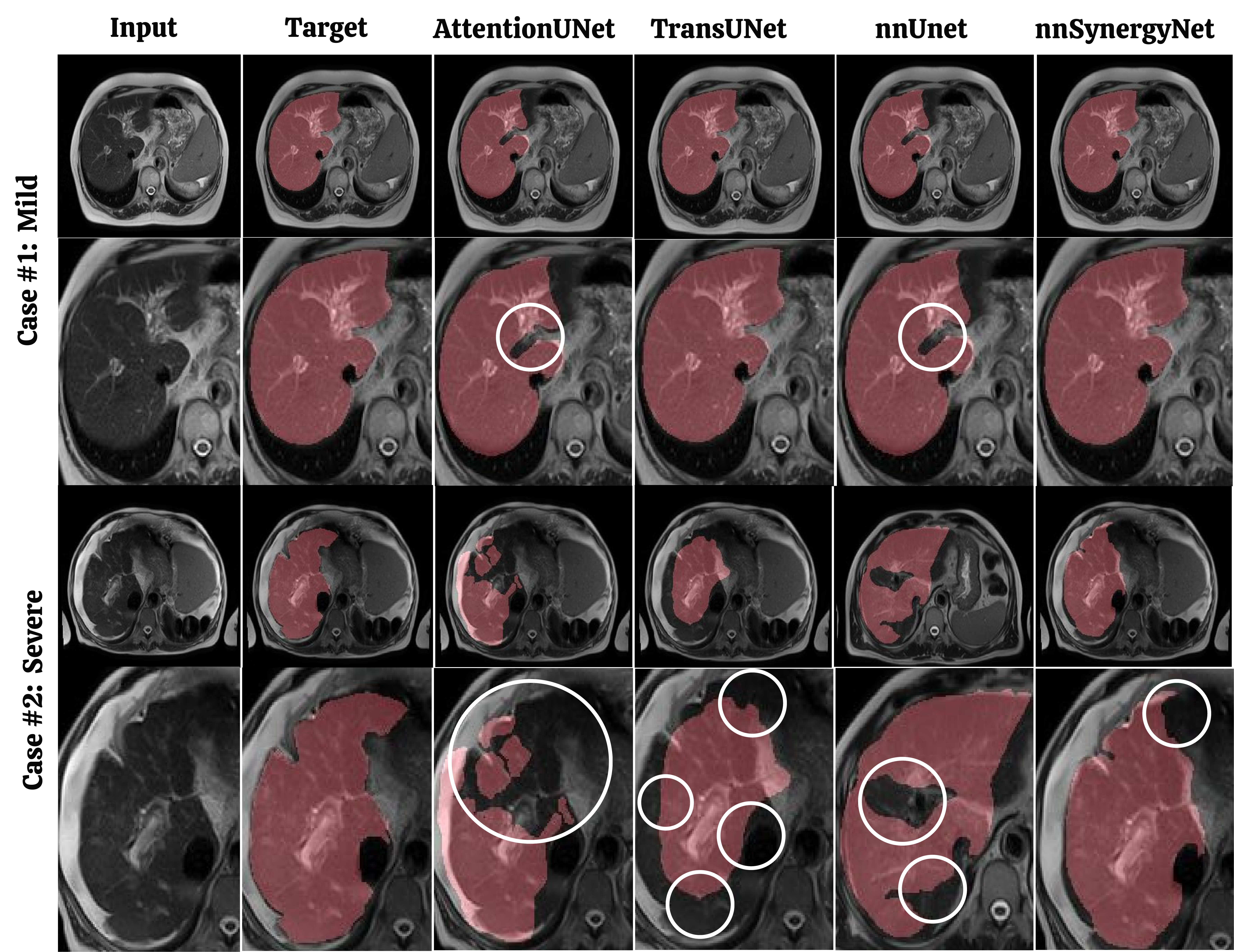}
    \caption{Qualitative results of different models on segmenting mild and severe cirrhosis from abdominal  T2W MRI scans. The white bounding circles show major errors made by the models.}
    \label{qualitative-t2}
\end{figure*}

\begin{figure*}[!t]
    \centering
\includegraphics[width=0.8\textwidth]{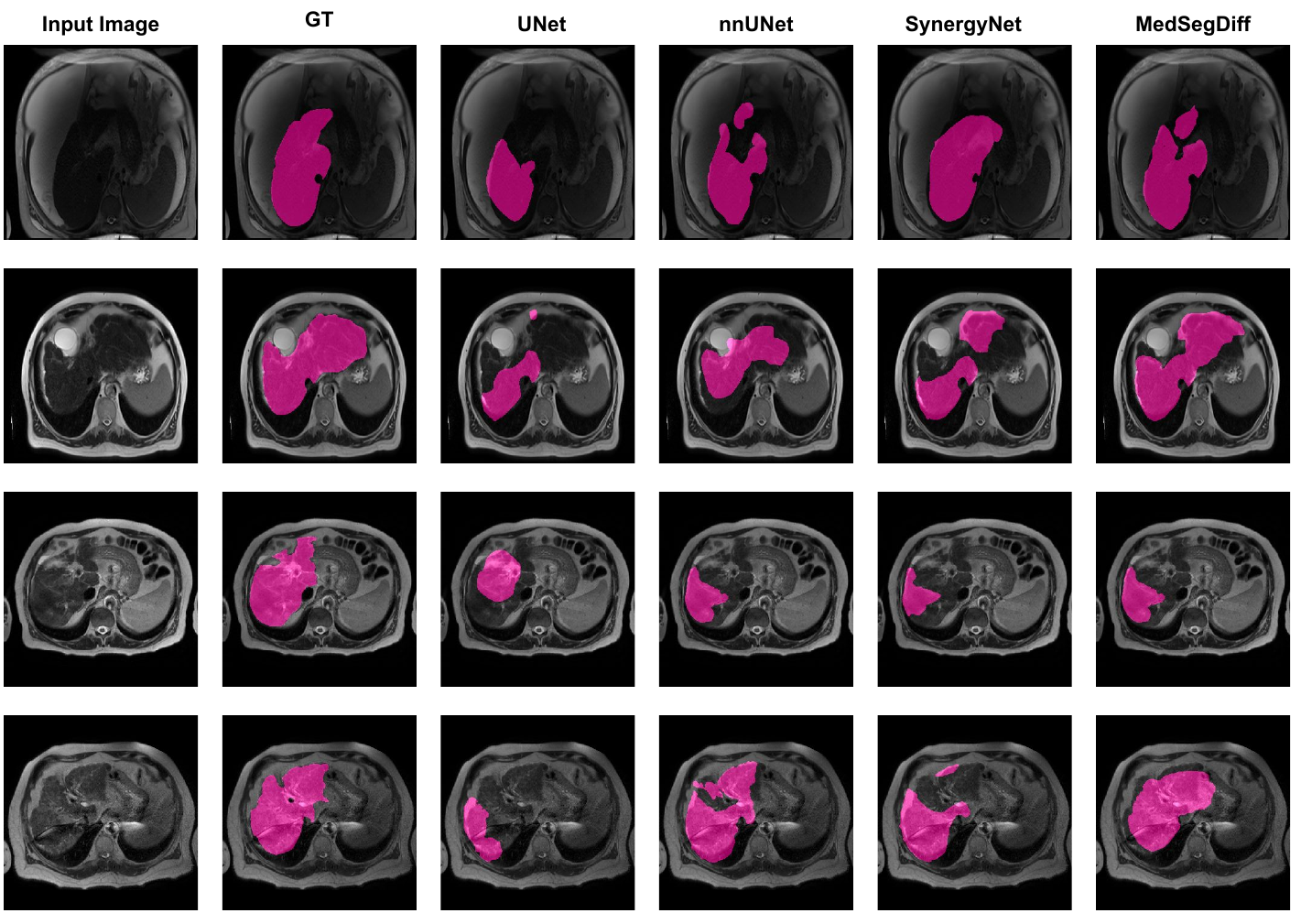}
    \caption{The figure shows the qualitative results examples of different models on segmenting mild and severe cirrhosis from abdominal T2W MRI scans. From the figure, it can be observed that MedSegDiff is the best choice.}
    \label{fig:enter-label2D}
\end{figure*}

\textbf{Qualitative Validation of Segmentation Masks on both modalities:} 
Fig.~\ref{qualitative-t1} and Fig.~\ref{qualitative-t2} show qualitative results for T1W and T2W samples, respectively. These results demonstrated that existing models perform well in segmenting cirrhotic liver under mild conditions. As highlighted by white boundaries, these models suffer under moderate-to-severe cases due to the poor texture of MRIs caused by cirrhosis scarring. \textit{nnSynergyNet3D} consistently performed well even for advanced cirrhotic livers compared to other SOTAs such as \textit{nnUNet, TransUNet}, and \textit{Attention-UNet}. This is likely because of its auto-configured, hybrid, and synergistic nature, enabling it to capture the texture of the cirrhotic liver more effectively. Figure~\ref{fig:enter-label2D} shows the qualitative results of different methods on \textbf{CirrMRI600+$\rightarrow$T2W} datasets. From the qualitative results, it can be observed that models such as UNet produce under-segmentation, whereas nnUNet produces over-segmentation. SynergyNet also produces shows under and over-segmentation for diverse cases. However, MedSegDiff is better at handling complexity.  The team of four clinicians verified all the segmented volumetric scans. We obtained high inter-observer agreement (kappa scores of 0.89 for T1W and 0.87 for T2W).

\begin{table*}[!t]
    \caption{Comparative benchmark of SOTA 3D segmentation networks on \textbf{CirrMRI600+ $\rightarrow$T2W}. Bold shows the best performance while red is the second-best.} 
    \centering
    \footnotesize
    \begin{adjustbox}{max width=\textwidth}
\begin{tabular}{lccccccc}
    \toprule
\textbf{Method}	&  \textbf{mIoU}	& \textbf{Dice}& \textbf{HD95} & \textbf{Precision} & \textbf{Recall}	& \textbf{ASDD}  &  \\
\midrule    

\texttt{VNet}~[\cite{milletari2016v}]	& 68.98	&70.01	&35.67	&69.98	&70.56	&7.18 \\

\texttt{Attention UNet}~[\cite{oktay2018attention}] & 68.72 & 79.18 & 37.87 & 79.99 &83.21    &  7.53 \\

\texttt{SynergyVNet3D}~[\cite{gorade2024synergynet}] & 75.17 & 77.56 &	28.19 &	83.78 &	85.42 &	5.79 \\

\texttt{TransBTS}~[\cite{wenxuan2021transbts}] &62.80  &74.88 &43.73 &  76.69   & 79.75  & 8.18   \\ 
\texttt{UXNet3D}~[\cite{lee20223d}] & 72.11  & 82.16 & 32.01 &84.83 &83.68  & 6.00 \\ 

\texttt{TransUNet3D}~[\cite{chen2021transunet}] 	&77.89	&79.09	&34.11	&78.11	& 79.97	&6.69 \\

\texttt{LinTransUnet}~[\cite{zhang2022dynamic}]	&80.08	&82.11 &26.01	&84.21	&86.17	&5.98\\

\texttt{SwinUNeTr}~[\cite{hatamizadeh2021swin}]	&79.89	&81.21	&32.78	&80.05	&81.10	&6.19\\

\texttt{nnUNet3D}~[\cite{isensee2021nnu}] 	&82.11	&84.76	&27.73	&\textcolor{red}{85.78}	&86.66	&4.76\\

\texttt{nnFormer3D}~[\cite{zhou2023nnformer}]	&\textbf{83.42}	&\textcolor{red}{86.47}	&\textcolor{red}{25.92}	&\textbf{87.67}	&\textbf{88.02}	&\textcolor{red}{4.04}\\

\texttt{nnSynergyNet3D}~[\cite{gorade2024synergynet}]	&\textcolor{red}{83.01}	&\textbf{86.51}	&\textbf{24.19}	&85.66	&\textcolor{red}{87.01}	&\textbf{3.96} \\

\bottomrule
\end{tabular}
\end{adjustbox}
\label{table:T2segmentation_methods}
\end{table*}

\begin{table*}[!h]
    \centering
    \footnotesize
        \caption{Comparative benchmark of SOTA 2D segmentation networks on \textbf{CirrMRI600+ →T2W 2D}.\textbf{ Bold} shows the best performance, and \textcolor{red}{red} shows  the second-best performance.}
    \begin{adjustbox}{max width=\textwidth}
  
        \begin{tabular}{l|c|c|c|c|c|c}
            \toprule
            \textbf{Method} & \textbf{mIoU} & \textbf{Dice} & \textbf{HD95} & \textbf{Precision} & \textbf{Recall} & \textbf{ASSD} \\
            \midrule
            UNet   &  0.6772  & 0.6900  & 38.22 &  0.7112  & 0.7592 &  10.11 \\
            AttentionUNet &0.7089  & 0.7288 & 36.19 & 0.7377 & 0.7689  & 9.28  \\
            nnUnet-2D    & 0.7229 & 0.7418 & 34.56 & 0.7662    & 0.7999 & 8.78  \\
            Trasunet     & 0.7219 & 0.7457 & 31.11 & 0.7447    & 0.7812 & 8.66  \\
            Synergynet   & \textcolor{red}{0.7383} & \textcolor{red}{0.7592} & \textcolor{red}{30.94} & {\textbf{0.7882}}    & {\textbf{0.8222}} & \textcolor{red}{7.55}  \\
            MedSegDiff   & \textbf{0.7489} & \textbf{0.7667} & \textbf{30.89} & \textcolor{red}{0.7789}    & \textcolor{red}{0.8192} & \textbf{7.34} \\
            \bottomrule
        \end{tabular}
    \end{adjustbox}
    \label{tab:t2w_2d}
\end{table*}

\textbf{Consistency Analysis of CirrMRI600+ Masks with Benchmarked Methods:}
We established six baseline models for our \textbf{CirrMRI600+$\rightarrow$T2W 2D} dataset, spanning diverse architectures: UNet, AttentionUNet, nnUNet2D (with deep supervision), TransUNet (transformer-based), SynergyNet, and MedSegDiff (diffusion-based) (See Table~\ref{tab:t2w_2d}). This range covers key approaches in medical image segmentation. Quantitative comparisons revealed SynergyNet as the top UNet-based performer, achieving a mIoU of 0.7383, Dice coefficient of 0.7592, HD95 of 30.94, precision of 0.7882, recall of 0.8222, and ASDD of 7.55. However, MedSegDiff outperformed all methods with a mIoU of 0.7489, Dice coefficient of 0.7667, HD95 of 30.89, and ASDD of 7.34, while SynergyNet maintained higher precision and recall values.


\textbf{Technical Challenge-Single Center Study:} While CirrMRI600+ offers a significant leap forward in liver cirrhosis segmentation research, it's important to acknowledge its limitations. One such limitation is its single-center nature. Ideally, a truly comprehensive dataset would encompass data from multiple medical institutions, capturing the potential variations in acquisition protocols and patient populations encountered in real-world clinical settings. However, CirrMRI600+ addresses this limitation by prioritizing other key strengths. The dataset was very carefully curated, ensuring the highest quality annotations and ground truth segmentation for all included scans. Furthermore, CirrMRI600+ boasts a rich heterogeneity in disease states within the cirrhotic population. By carefully selecting a diverse range of cirrhotic liver presentations, the dataset ensures generalizability to a broader spectrum of cirrhosis cases. 

\textbf{Technical Challenge-Not all T1Ws are contrast-enhanced:} Another weakness was that more than 95\% of the T1W images were contrast enhanced and not all possible contrast types and phases of contrast administration were represented in the dataset. Although the cross-modality testing shows strong generalizability, small annotations might have been missed, and our annotations might not be correct pixel precise for all the cases. Other modalities, such as computed tomography (CT) scan, ultrasound, elastography, and magnetic resonance elastography, were absent besides MRI. 

\textbf{Vision for Multi-Organ Annotations:} To ensure high-quality annotations while maintaining efficiency, CirrMRI600+ leverages a semi-automated approach in ground truth construction. Deep learning model predictions served as a starting point, followed by refinement from radiologists. This approach enabled radiologists to focus on potential algorithm failures and refine the output masks more efficiently, saving valuable annotation time and cost. Rigorous benchmarking using the \textit{nnSynergyNet3D} method on both T1W and T2W scans demonstrates promising performance, achieving Dice Similarity Coefficients (DSC) of 87.89\% and 86.51\%, respectively. CirrMRI600+ lays the foundation for a future multi-organ dataset. Our long-term vision is to expand the annotations to include additional organs such as kidneys, spleen, pancreas, and major vessels, creating a comprehensive resource for abdominal organ segmentation tasks for multimodal MRIs.

\textbf{Other Technical Challenges and Potential Solutions:} While our benchmark results demonstrate strong performance for liver segmentation (achieving DSC of 87.89\% for T1W and 86.51\% for T2W with nnSynergyNet3D), significant challenges and research opportunities remain in the domain of cirrhotic liver analysis. We identify several key areas for future investigation using the CirrMRI600+ dataset:

\textbf{Segmentation:} Despite the promising results, our qualitative analysis reveals persistent difficulties in accurately delineating certain anatomical regions, particularly in cases with severe cirrhosis. These challenges include:

\textit{(1) Boundary ambiguity in advanced cirrhosis:} The irregular, nodular boundaries characteristic of advanced cirrhotic livers remain difficult to precisely segment, especially where the liver interfaces with adjacent structures having similar intensity profiles.

\textit{(2) Segmentation of heterogeneous parenchyma:} The variable signal intensities within severely fibrotic regions often lead to under-segmentation of areas that deviate significantly from typical liver appearance.

\textit{(3) Handling of focal lesions:} The presence of focal lesions (including regenerative nodules and potential hepatocellular carcinoma) introduces additional complexity that current methods struggle to consistently address.

\textit{(4) Inter-modality performance gap:} The observed differences in segmentation performance between T1W and T2W sequences suggest opportunities for developing specialized approaches that better leverage the unique characteristics of each modality.

\section*{Usage Notes}
The entire dataset can be downloaded from the OSF repository~\cite{OSF}. To process the provided images and segmentation maps (for all the compared benchmarking methods), we highly recommended using medical imaging tools such as ITKSnap, 3D-Slicer, Amira, or similar dicom viewers. We verified that all NIfTI files (segmentations and images) and DICOM files can be loaded correctly with 3D-Slicer (https://www.Slicer.org) and ITKSnap.

\section*{Code Availability}
All the segmentation codes, model files, and customized instructions are available to be downloaded from our dedicated GitHub repository (https://github.com/NUBagciLab/CirrMRI600Plus)~\cite{CirrMRI600Plus}.


\section*{Acknowledgements}
This project is supported by NIH funding: R01-CA246704, R01-CA240639, U01-DK127384-02S1, and U01-CA268808.   

\section*{Author contributions statement}
A. Medetalibeyoglu, G. Durak, D. Jha and U. Bagci conceptualized the work. U. Bagci, G. Durak and A. Medetalibeyoglu were critical in designing the study and facilitating the data collection. D. Jha, O. K. Susladkar and V. Gorade performed the baseline experiments. G. Durak led the data annotation process coordinating effort with a team including  E. Keles, M. Antalek, D. Seyithanoglu, T. Cebeci, H. Ertugrul Aktas, G. D. Kartal, S. Kaymakoglu, S. M. Erturk, Y. Velichko, D. Ladner, A. A. Borhani, and A. Medetalibeyoglu. All the images were reviewed by the radiology team first, and consensus was achieved with the clinical team about the annotation style, time, repetition, and other clinical questions related to cirrhosis severity estimation. T. Cebeci, G. Kartal, S. Kaymakoglu, and S. Erturk helped in data collection, IRB, and secure transfer of the files. All authors contributed to the paper's writing and revision.

\subsection*{Code statement}
The code for baseline methods are available at \url{https://github.com/NUBagciLab/CirrMRI600Plus}.

\subsection*{Competing interests}
The authors declare no competing interests regarding this publication and the dataset.

\end{document}